\documentclass[twocolumn,prl,aps]{revtex4}
\usepackage{psfig}
\newcommand{\be}{\begin{equation}}
\newcommand{\ee}{\end{equation}}
\newcommand{\bea}{\begin{eqnarray}}
\newcommand{\eea}{\end{eqnarray}}

\begin{document}
\title{Rapidity dependence of large-$p_\perp$ hadron production at RHIC}
\author{Alberto Polleri and Feng Yuan}
\affiliation{Institut f\"ur Theoretische Physik, Universit\"at Heidelberg, 
             Philosophenweg 19, D-69120 Heidelberg, Germany}


\begin{abstract}
We study the dependence of parton energy loss (quenching) on rapidity in 
ultra-relativistic nuclear collisions at RHIC.
This can provides invaluable information on the density of the medium, 
which should be more dilute going away from mid-rapidity, 
thereby reducing the effect of quenching. 
We predict a clear effect at moderate transverse momenta $\sim$ 3 GeV.
\end{abstract}


\maketitle

Energy loss of high energy quark and gluon jets in the dense matter produced
in ultra-relativistic heavy ion collisions is expected to lead to jet quenching 
and can potentially probe the quark-gluon plasma formed in these reactions
\cite{gpw,gw94,Baier,Zakharov,Wiedemann,glv,Wangd}. 
Recent preliminary data from the first run of RHIC at BNL
suggest the existence of this phenomenon \cite{recent}, where moderate 
$p_T \sim 2-6$ GeV hadron spectra were found to be suppressed by 
a factor $\sim 2-5$ relative to the pQCD expectations.
In addition, the suddenly saturated azimuthal asymmetric moment $v_2(p_T)$
for values of $p_T > 2$ GeV was interpreted as evidence of jet quenching in a 
dense gluon plasma with rapidity density of the initially produced gluons
in the range of $dN^g/dy\sim 500-1000$ \cite{v2-pt}.
However, to finally constrain the amount of jet quenching which affects high
momentum hadrons at RHIC, we must seek further independent observables to 
test these phenomena and support their existence.

Previous studies mostly focus on the suppression of $p_T$
spectra in nucleus-nucleus collisions compared with the
proton-proton case at the same energy, or in central nucleus-nucleus
collisions compared with the peripheral ones. This is always done at mid-rapidity.
The point we want to stress is that energy loss effects depend on the density of
the environment, {\it i.e.}, higher medium densities mean higher amount of energy
loss. Although one can study the dependence on the medium density by varying the
impact parameter, at fixed impact parameter $b$ for nucleus-nucleus collisions,
also different rapidities provide different densities of the
medium, through which the high momentum jet travels.
As illustrated in Fig.~1, when a high energy jet travels through 
the quark-gluon plasma produced at the early stage of nucleus-nucleus collisions,
the medium density is much larger at mid-rapidity than in the forward or 
backward directions (large rapidity). 
Because jet energy loss is proportional to the density of the
local medium\cite{gw94,Baier,Zakharov,Wiedemann,glv,Wangd,v2-pt}, 
at different rapidities the energy loss of the fixed $p_T$ jet will be
different, and the large $p_T$ hadron spectrum from jet fragmentation will
also have different behavior.
Therefore, by scanning the rapidity of hadrons produced in central 
nucleus-nucleus collisions, one can observe the variation of energy loss effects 
and, as we will show with the following calculations, its influence on the 
rapidity spectrum of charged hadrons at fixed $p_T$. A study based on an
analogous idea was recently made for $J/\psi$ suppression as function 
of $x_F$ \cite{psi}.

\begin{figure}[t]
\centerline{\psfig{figure=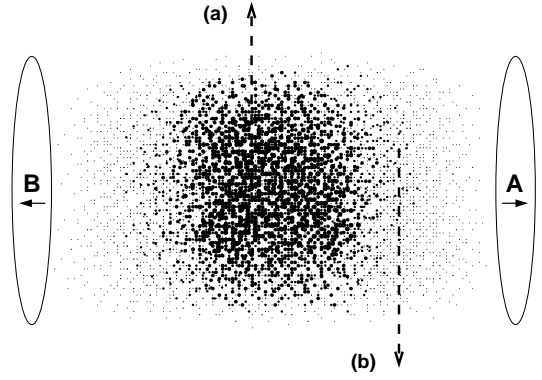,width=7cm}}
\protect\caption{Illustration of the spatial extention a nucleus-nucleus 
collision at fixed time in the c.m. frame. Dashed lines show trajectories
of jets produced with small (a) and large (b) values of $y$, traversing dense
and dilute parts of the medium, respectively.} 
\label{cartoon} 
\end{figure}

Large $p_T$ hadron production in $pp$ collisions can be expressed in a standard
pQCD form as
\bea
\frac{d\sigma^h_{pp}}{dyd^2p_T} & = & K\int dx_adx_bd^2k_ad^2k_b
\ g_p(k_a)\ g_p(k_b) \label{prot} \\
& & \!\!\!\!\!\!\!\!\!\!\!\!\!\!\!\!\!\!\!\!\! \times 
\ f_{a/p}(x_a,Q^2)\ f_{b/p}(x_b,Q^2)
\ \frac{d\hat\sigma_{ab \rightarrow cX}}{d\hat t}
\ \frac{D_{h/c}(z_c,Q^2)}{\pi z_c}\,, \nonumber
\eea
where $f_{i/p}$ is the distribution function of parton $i$ in the proton, 
$d\hat\sigma/d\hat t$ are parton-parton coss sections and $D_{h/c}$ is the 
fragmentation function of parton $c$ in hadron $h$. We include
intrinsic transverse momentum effects adopting the phenomenological $k_T$-kick 
model with the Gaussian distribution for the intrinsic $k_T$ of a parton inside 
the proton
\begin{equation}
\label{intrinsic}
g_p(k_T) = (\pi \Delta_p^2)^{-1} \exp(- k_T^2 / \Delta_p^2).
\end{equation}
In the numerical calculations, we choose the GRV94 parameterization for 
the parton distribution functions \cite{grv94}, 
and the KKP set for the fragmentation functions into charged hadrons \cite{kkp}.
With the scale choice of $Q^2=p_c^2$, and an overall factor $K=1.5$, 
we can reproduce charged hadron spectra in $pp$ collisions
in various experiments and also the UA1 data for $p\bar p$ collisions
at $\sqrt{s}=200$ GeV. There we use the energy dependent values for 
$\Delta_p^2$ \cite{papp} ($\Delta_p^2=1.0$ GeV$^2$ at $\sqrt{s}=200$ GeV).

We now turn to the case of proton-nucleus collisions. The first feature we 
need to discuss is that of parton shadowing. 
Although some approaches attempt to solve the problem in general terms, 
a complete solution is not yet at hand. Since this aspect is not our main 
focus, we make use of the popular EKS98 parameterization \cite{eks98} 
and modify it to take into account impact parameter 
dependence. This leads to the factor  
\be
R_{i/A}(x,Q^2,b) = 1 - \frac{T_A(b)}{\langle T_A \rangle}\ S_{i/A}(x,Q^2)\,,
\ee
where $\langle T_A \rangle = A^{-1}\!\int d^2b\ T_A^2(b)$ is the mean value of the 
thickness function and
$S_{i/A}(x,Q^2)$ is the shadowing factor of the EKS parameterization.
In this way we satisfy the relation
\bea
\langle\,R_{i/A}\,\rangle (x,Q^2) & = & A^{-1}\int d^2b\ T_A(b)\ R_{i/A}(x,Q^2,b)
\nonumber \\
& = & 1 - S_{i/A}(x,Q^2)\,
\eea
which reproduces, by construction, the EKS98 parton distribution ratios.

Another feature which appears in $pA$ collisions is the broadening of transverse 
momentum spectra (Cronin effect). This is understood as parton rescattering 
during the traversal of the projectile through the nucleus. 
This effect is parameterized by
convoluting the spectra with a Gaussian function whose width
is proportional to the mean number of rescatterings the projectile suffers inside
the nucleus \cite{Wang1,levai}.
With this method, the $p_T$ broadening effects on hadron production 
in $pA$ and $AB$ collisions 
have been previously studied with a modification of the intrinsic $k_T$ 
distribution of a nucleon by convoluting it 
with the broadening factor \cite{Wang1,levai}.
Therefore, the initial intrinsic $k_T$ distribution of partons will be modified
with respect to Eq.~(\ref{intrinsic}) to the form
\begin{equation}
g_A^{in}(k_T,b) = (\pi \Delta_A^2(b))^{-1} \exp(- k_T^2 / \Delta_A^2(b))\,,
\end{equation}
with 
\be
\Delta_A^2(b) = \Delta_p^2 + \Delta_{in}^2(b)\,,\ \ \ \ 
\Delta_{in}^2(b) = \lambda \sigma_{NN}T_A(b)\,,
\ee
where $\lambda$ is a coefficient corresponding to the average value of the
additional squared momentum, acquired in each collision by parton $i$, and 
$\sigma_{NN}T_A(b)$ is the number of scattering the projectile nucleon suffers
inside the nucleus.
In our calculations, we also take into account the broadening effects
to the final jet spectrum prior to fragmentation and convolute it with
the broadening distribution
\begin{equation}
g_A^{fin}(k_T,b) = (\pi \Delta_{fin}^2(b))^{-1} \exp(- k_T^2 / \Delta_{fin}^2(b)).
\end{equation}
with $\Delta_{fin}^2(b) = \Delta_{in}^2(b)$.
In the following we differentiate the broadening effects between
quarks and gluons, where the quark one is $4/9$ smaller than the
gluon one.
By fitting the charged hadron spectrum from various $pA$ data,
we obtained the coefficient $\lambda=0.1$ GeV$^2$, which roughly agrees
with previous studies \cite{Wang1,levai}.

Considering now the $AB$ case, we first of all generalize the nuclear effects 
as discussed above. This is straightforward for shadowing and initial broadening.
In fact each colliding nucleus is affected and not only the target as in the $pA$
case. Concerning final state
broadening we must now take into account both nuclei at once, therefore
$\Delta_{fin}^2 = \lambda \sigma_{NN}(T_A+T_B)$. Then we come to discuss the 
key issue of the present paper, that is parton energy loss. 
In the literature, there have been intense studies on this subject
\cite{gw94,Baier,Zakharov,Wiedemann,glv,Wangd}. 
These studies show that the energy loss of the fast parton in 
a dense medium depends on the medium density, 
the initial parton energy and the path length of
the parton traversing the medium.
In different kinematical regions, the total energy loss has a different 
functional dependence on the above mentioned factors \cite{review}.
However, in the region of interest for the production of few GeV hadrons, 
it is widely believed that the
energy loss is proportional to the parton density and to the square of the 
path length(in a static medium).
Then, we make a simple observation. 
If a jet is produced at mid-rapidity, it will 
encounter a dense medium, while at large enough rapidities the
energy loss effect disappears. We therefore assume that a parton $c$ 
loses energy because of gluon radiation induced by the medium, in the amount
\be
\label{deltae}
\Delta E_c(y,E_c) = q_c\ E_c\ \frac{dn}{dy}\,,
\ee
where $q_c$ is the average value of energy loss which implicitly also includes
the size of the medium. Its value has to be determined and for quarks it is 
$4/9$ that of gluons.
In the simple parameterization above, we assume that the energy loss is 
also proportional to the original jet energy, according to the recent 
findings within the GLV formalism \cite{glv,levai1}.
We note that in the literature, there exist various
formalisms which give rise to a different energy dependence of the parton 
energy loss \cite{glv,v2-pt,review}. 
In the GLV formalism, the linear energy dependence of jet energy loss adopted 
here is mainly caused by the kinematic limit of the gluon radiation at low 
energy \cite{glv}.
This linear dependence may not be valid at large energy. In fact, at 
asymptotically high energies, the dependence becomes 
logarithmic \cite{glv,v2-pt}.
Moreover, a recent study on this subject \cite{Wangd} shows that 
thermal gluon absorption can reduce the
energy loss for a low energy jet and change the energy dependence
of the energy loss to stronger than linear at low energy.
Nevertheless, in our calculations we concentrate on the production of few GeV 
hadrons where the linear dependence for jet energy loss should be a reasonable
approximation \cite{glv,v2-pt,levai1}.

Eq.~(\ref{deltae}) also contains the rapidity dependence of jet energy loss, 
which relies on the density of the medium $dn/dy$. We note 
that at large rapidities the time evolution of the medium may be quite 
different from that at mid-rapidity ($y=0$), and jet energy loss 
might have a complicated dependence on geometry.
However, we argue that at larger rapidity, the dominant effects to the jet 
energy loss is caused by the lower medium density.
This means less energy loss at larger rapidities. 
So, as a first estimate of the rapidity dependence, in our calculations we 
made the simple but intuitive assumption contained in Eq.~(\ref{deltae}) to 
parameterize the rapidity dependence of jet energy loss.
For the medium density $dn/dy$, we further assume for simplicity that the plasma 
density at the early stage is proportional to the 
measured rapidity density of charged hadrons, therefore including both soft and
hard produced particles. 
The shape was fitted as in \cite{psi} to the recent PHOBOS data and here 
normalized to 1 at $y = 0$. We do not take into explicit account the time
evolution of the plasma which, due to geometric and dynamical effects, requires
a more detailed study of its properties. We therefore consider only the pre-factor 
$q_c$ which contains the average effect of the space-time evolution.

If a parton loses energy, its fragmentation is then modified
\cite{Wangh,GuoWang}, since its momentum fraction which is converted into 
a hadron is reduced. This can be approximately realized by 
introducing the momentum fraction $z_c' = p_\perp/(p_c - \Delta E_c(p_c))$, 
being $p_c = p_\perp/z_c$ (Although see \cite{bdmsnew}), and changing the 
fragmentation function as
\be
D_{h/c}(z_c,Q^2) \rightarrow 
 \frac{z_c'}{z_c}\ D_{c/h}(z_c',Q^2)\,.
\ee
In this way we neglect a small contribution due to the fragmentation of
radiated soft gluons.

Substituting all the elements which appear in describing nucleus-nucleus 
collisions in addition
to the $pp$ case, we obtain the final form for the hadron spectrum
\bea
\frac{dN_{AB}^h}{dyd^2p_T}(b) &=& K\int d^2b_A\ dx_adx_bd^2k_ad^2k_b\ d^2k_c\label{nuc} \\
& & \!\!\!\!\!\!\times\ g_A^{in}(k_a,b_A)\ R_{a/A}(x_a,Q^2,b_A)\ f_{a/p}(x_a,Q^2) 
\nonumber \\
& & \!\!\!\!\!\!\times\ g_B^{in}(k_b,b_B)\ R_{b/B}(x_b,Q^2,b_B)\ f_{b/p}(x_b,Q^2)
\nonumber \\
& & \!\!\!\!\!\!\!\!\!\!\!\!\!\!\!\!\!\!\!\!\!\!\!\!\!\!\!\!\!\!\! \times 
\ \frac{d\hat\sigma_{ab \rightarrow cX}}{d\hat t}
\ g_{AB}^{fin}(\vec k_c\!-\!\vec p_T/z_c,b_A,b_B)
\ \frac{z_c'\ D_{h/c}(z_c',Q_f^2)}{\pi z_c^2}\,. \nonumber
\eea
In this way we have completed the discussion of the technical part of the 
paper and proceed to examine the results of the computations and their 
physical implications.

\begin{figure}[htb]
\centerline{\psfig{figure=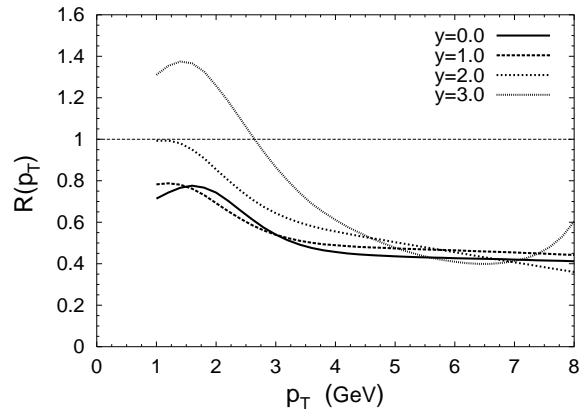,width=8cm}}
\protect\caption{Transverse momentum dependence of the ratio of spectra in
$AB$ collisions with respect to the $pp$ case, for different rapidity values.}
\end{figure}
\begin{figure}[htb]
\centerline{\psfig{figure=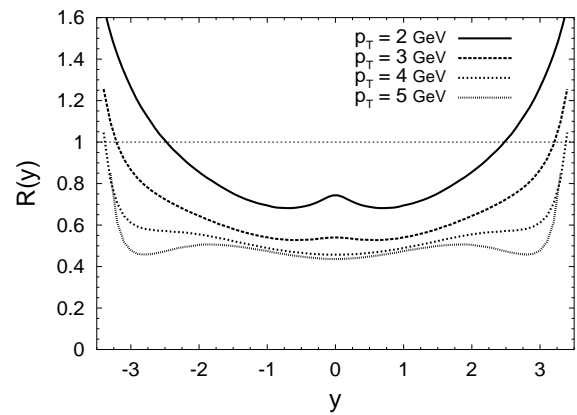,width=8cm}}
\protect\caption{Rapidity dependence of the ratio of spectra in
$AB$ collisions with respect to the $pp$ case, for different transverse
momentum values.}
\end{figure}

Our numerical results are presented in Figs.~2-3, where we display ratios
of spectra in nucleus-nucleus collisions respect to $pp$ collisions. Actually
we computed 
\be
R^h_{AB}(y,p_T,b) = \frac{\mbox{\Large $\frac{dN^h_{AB}}{dyd^2p_T}$}(b)}
{T_{AB}(b)\, \mbox{\Large $\frac{d\sigma^h_{pp}}{dyd^2p_T}$}}\,,
\ee
where $T_{AB}(b)$ is the usual overlap function, and took $b=0$.
Fig.~2 is for the ratios of $p_T$ spectra at different rapidities, and 
Fig.~3 for the ratios of rapidity spectra at different $p_T$ values.
All the results are for $Au+Au$ collisions at $\sqrt{s}=200$ GeV,
relevant to RHIC.
For the energy loss coefficient we use $q_c = 0.27$ 
with which we can reproduce the suppression of the charged hadron 
spectrum in central collisions observed at $\sqrt{s}=130$ GeV.
In Fig.~2, we see that for different rapidities the ratios have
different behaviors. In general, at larger $y$, the suppression is reduced,
as expected, due to the decrease in the medium density.
Especially, at very large rapidity one can really observe
the Cronin effect. The rise at $p_T > 7$ GeV of the curve for $y=3$ arises due 
to the vicinity of the kinematical boundary for hadron production, where our 
treatment of final state $p_T$ broadening as folding with Gaussians 
is no more valid.

Even more interesting results can be found for the ratios as functions of rapidity
at different values of $p_T$, as illustrated in Fig.~3. 
At mid-rapidity, because of the high density of the medium, the largest
energy loss produces the strongest suppression. As rapidity increases, the 
energy loss effect dies out and the ratio increases. 
Especially for the largest values of
$p_T$ what is clear is the presence of a concavity around mid-rapidity in the
ratio. This feature should not be present in absence of jet quenching, while 
our calculations provide it clearly. This is the result of our investigation and
we argue that it should be a good signal to observe energy loss at work.

The idea of scanning the medium by computing rapidity spectra can be extended to 
the study of the $y$ dependence of the $v_2$ coefficient of azimuthal asymmetry. 
We expect its value to vary significantly with rapidity. Moreover, a comprehensive
study of direct photon production could be a cross check of the present work, since 
photon should not experience energy loss end therefore their spectra should not
be quenched. 

In conclusion, in this paper we have calculated the rapidity dependence
of hadron spectra for high values of $p_T$ in $Au+Au$ collisions at 
$\sqrt{s}=200$ GeV. Comparing the results to the $pp$ case, we found a strong
signal to test the effect of jet energy loss. The observation of such a behavior
within future RHIC data will give a new test and support for the idea of
jet quenching, therefore providing important information about the existence and
properties of the quark-gluon plasma.

\noindent {\bf Acknowledgments}:  The authors would like to thank J.~H\"ufner, 
B.~Kopeliovich, H.J.~Pirner, U.~Wiedemann, and X.~N.~Wang for several 
stimulating discussions. 
This work has been partially supported by the German Federal Ministry BMBF 
with grant No.~06 HD 954.

\end{document}